\documentclass[%
reprint, twocolumn,
superscriptaddress,
showpacs,preprintnumbers,
nofootinbib,
amsmath,amssymb, aps, pra,
]{revtex4}

\usepackage{graphicx}
\usepackage{dcolumn}
\usepackage{bm}
\usepackage{color} 
\usepackage{CJK}

\def\ra{{\rangle}}

\newcommand{\beq}{\begin{equation}}
\newcommand{\eeq}{\end{equation}}
\newcommand{\beqa}{\begin{eqnarray}}
\newcommand{\eeqa}{\end{eqnarray}}

\DeclareMathOperator{\arccot}{arccot}

\begin{document}
\title{Shortcuts to adiabaticity in three-level systems using Lie transforms}
\author{S. Mart\'\i nez-Garaot}
\affiliation{Departamento de Qu\'{\i}mica F\'{\i}sica, UPV/EHU, Apdo
644, 48080 Bilbao, Spain}
\author{E.  Torrontegui}
\affiliation{Departamento de Qu\'{\i}mica F\'{\i}sica, UPV/EHU, Apdo
644, 48080 Bilbao, Spain}
\affiliation{Institute of Chemistry, The Hebrew University, Jerusalem 91904, Israel}
\author{Xi Chen}
\affiliation{Department of Physics, Shanghai University, 200444
Shanghai, People's Republic of China}
\author{J. G. Muga}
\affiliation{Departamento de Qu\'{\i}mica F\'{\i}sica, UPV/EHU, Apdo
644, 48080 Bilbao, Spain}
\affiliation{Department of Physics, Shanghai University, 200444
Shanghai, People's Republic of China}
\date{\today}
\begin{abstract}
Sped-up protocols (shortcuts to adiabaticity) that drive a system quickly to the same populations than a slow adiabatic process 
may involve Hamiltonian terms difficult to realize in practice. We use the dynamical symmetry of the Hamiltonian to 
find, by means of Lie transforms, alternative Hamiltonians  that achieve the same goals without the problematic terms.
We apply this technique to three-level systems 
(two interacting bosons in a double well, and  beam splitters with two and three output 
channels) driven by Hamiltonians that belong  
to the  four-dimensional algebra U3S3. 
\end{abstract}
%
%
\maketitle
%
\section{Introduction}
``Shortcuts to adiabaticity'' are  manipulation protocols 
that take the system quickly to the same populations, or even the same state, 
than a slow adiabatic process  \cite{review}.  
Adiabaticity is ubiquitous to prepare the system state in atomic, molecular and optical physics, so many applications of this concept 
have been worked out, both in theory and experiment \cite{review}.    
Some of the engineered Hamiltonians that speed up the adiabatic process in principle, may involve 
terms difficult or impossible to realize in practice.  
In simple systems, such as single particles transported \cite{transport_2011} or expanded by harmonic potentials \cite{expansions_2010},
or  two-level systems \cite{2ls_2010, IP, 2ls_2011}, the dynamical symmetry of the Hamiltonian could be used to eliminate the problematic  terms and provide instead  feasible Hamiltonians.
In this paper we extend this program to three-level systems whose Hamiltonians belong to a four-dimensional dynamical algebra.   
This research was motivated by a recent observation by Opatrn\'y and M\o lmer \cite{Molmer}. 
Among other systems they considered two (ultra cold) interacting bosons in a double well within a three-state approximation. 
Specifically the aim was to speed up a  
transition from a ``Mott-insulator'' state with one particle in each well, to a  delocalized ``superfluid'' state. 
The reference adiabatic process  consisted on slowly turning off the inter particle interaction while increasing the tunneling
rate. To speed up this process they applied a  
method to generate shortcuts based on adding a ``counterdiabatic'' (cd) term to the original time-dependent Hamiltonian
\cite{Rice,Berry09,2ls_2010},  but the evolution with the cd-term turns out to be difficult to realize in practice \cite{Molmer}.   
In this paper we shall use the symmetry of the 
Hamiltonian (its dynamical algebra) to find an alternative shortcut  
by means of a Lie transform, namely, a unitary operator in the Lie group associated with the Lie algebra.  
Since other physical systems have the same Hamiltonian structure the results are applicable to them too. 
Specifically the analogy between the time-dependent Schr\"odinger equation and the stationary wave equation 
for a waveguide in the
paraxial approximation \cite{Longhi_2008,Rev_Longhi,Rev_Nolte,Longhi_2011,Vitanov_2012, Tseng_2013} is used   
to design  short-length optical beam splitters with two and three output channels.

In Sec. \ref{the model} we describe the theoretical model for two indistinguishable particles in two wells. In Sec. \ref{CD} we summarize the counterdiabatic or transitionless tracking approach and apply it to the bosonic system. Sec. \ref{as} sets the approach based on unitary Lie transforms to produce alternative shortcuts. In Sec. \ref{is} we introduce the insulator-superfluid transition and  apply the shortcut designed in the previous section. In Sec. \ref{bs} we apply the technique to generate beam splitters with two and three output channels. Section \ref{discussion} discusses the results and open questions. Finally, in the Appendix \ref{algebra} some features of the Lie algebra of the system are discussed.   
\section{The model}
\label{the model}
An interacting boson gas in a two-site potential  is described within the Bose-Hubbard approximation \cite{Fisher, Zoller} by  
\beq
\label{H_BH}
H_0=\frac{U}{2}\sum^2_{j=1}n_j(n_j-1)-J(a_1a_2^\dag+a_1^\dag a_2),
\eeq
where $a_{j}$ ($a_{j}^\dag$) are the bosonic particle annihilation (creation) operators at the $j$-th site and $n_j$ is the occupation number operator. The on-site interaction energy is quantified by the parameter $U$ and the hopping energy by $J$. They are assumed to be controllable functions of time.  
For two particles the Hamiltonian in the occupation number basis $|2,0\rangle=\left ( \scriptsize{\begin{array} {rcccl} 1\\ 0\\0 \end{array}} \right)$, $|1,1\rangle=\left ( \scriptsize{\begin{array} {rcccl} 0\\ 1\\0 \end{array}} \right)$ and $|0,2\rangle=\left ( \scriptsize{\begin{array} {rcccl} 0\\ 0\\1 \end{array}} \right)$, is given by \cite{Molmer}
\beq
\label{H_0}
H_0=\left ( \begin{array}{ccc}
U & -\sqrt{2}J & 0\\
-\sqrt{2}J & 0 & -\sqrt{2}J \\
0 & -\sqrt{2}J & U
\end{array} \right)=UG_4-4JG_1,
\eeq
where
\beq
\label{G1_G4}
G_1=\frac{1}{2\sqrt{2}}\left ( \begin{array}{ccc}
0 & 1 & 0\\
1 & 0 & 1 \\
0 & 1 & 0
\end{array} \right), 
\, \, \,
G_4=\left ( \begin{array}{ccc}
1 & 0 & 0\\
0 & 0 & 0 \\
0 & 0 & 1
\end{array} \right).
\eeq
This Hamiltonian belongs to the vector space (Lie algebra) spanned by $G_1$, $G_4$, and 
two more generators, 
\beq
\label{generators}
G_2=\frac{1}{2\sqrt{2}}\left ( \begin{array}{ccc}
0 & -i & 0\\
i & 0 & i \\
0 & -i & 0
\end{array} \right), 
G_3=\frac{1}{4}\left ( \begin{array}{ccc}
1 & 0 & 1\\
0 & -2 & 0 \\
1 & 0 & 1
\end{array} \right), 
\eeq
with nonzero commutation relations
\begin{eqnarray}
\label{commutators}
[G_1, G_2]= iG_3,  
[G_2, G_3]= iG_1, 
[G_3, G_1]= iG_2,
\nonumber
\eeqa
\beqa
[G_4, G_1]= iG_2, 
[G_2, G_4]= iG_1.
\eeqa
This 4-dimensional Lie algebra, U3S3 \cite{Mac}, is described in more detail in the Appendix \ref{algebra}. 
To find the Hermitian basis we calculate $[G_1,G_4]$, and then all commutators of the result  
with previous elements. This operation is repeated for all operator pairs until no new, linearly independent operator appears.  

To diagonalize the Hamiltonian (\ref{H_0}) it is useful to  parameterize $U$ and $J$ as \cite{Molmer}
\beq
\label{parametrization_U}
U=E_0 \cos\varphi,\; 
J=\frac{E_0}{4} \sin \varphi,
\eeq
where $E_0=E_0(t)$ and $\varphi=\varphi(t)$, so that 
\beq
\label{H_0_2}
H_0=E_0 \left ( \begin{array}{ccc}
\cos\varphi & -\frac{1}{2\sqrt{2}} \sin \varphi & 0\\
-\frac{1}{2\sqrt{2}} \sin \varphi & 0 & -\frac{1}{2\sqrt{2}} \sin \varphi \\
0 &-\frac{1}{2\sqrt{2}} \sin \varphi & \cos\varphi
\end{array} \right).
\eeq
The instantaneous eigenvalues are
\beqa
\label{eigenvalues_1}
&&E_1=\frac{E_0}{2}(\cos\varphi-1),
\\
\label{eigenvalues_2}
&&E_2=E_0\cos \varphi,
\\
\label{eigenvalues_3}
&&E_3=\frac{E_0}{2}(\cos\varphi+1),
\eeqa
corresponding to the normalized eigenstates
\beqa
\label{eigenstates_1}
&&|\phi_1\rangle= \left ( \begin{array}{c}
\frac{1}{2}\sqrt{1-\cos\varphi}\\
\frac{1}{\sqrt{2}}\sqrt{1+\cos\varphi}\\
\frac{1}{2}\sqrt{1-\cos\varphi}\\
\end{array} \right), 
\\
\label{eigenstates_2}
&&|\phi_2\rangle=\frac{1}{\sqrt{2}} \left ( \begin{array}{c}
1\\
0\\
-1\\
\end{array} \right), 
\\
\label{eigenstates_3}
&&|\phi_3\rangle= \left ( \begin{array}{c}
\frac{1}{2}\sqrt{1+\cos\varphi}\\
-\frac{1}{\sqrt{2}}\sqrt{1-\cos\varphi}\\
\frac{1}{2}\sqrt{1+\cos\varphi}\\
\end{array} \right). 
\eeqa
\section{Counterdiabatic or transitionless tracking approach}
\label{CD}
For the transitionless driving or counterdiabatic approach  formulated by Demirplak and Rice \cite{Rice} or equivalently by Berry \cite{Berry09}, 
the starting point is a time-dependent reference Hamiltonian
\beq
\label{H_0_B}
H_0(t)=\sum_n|n_0(t)\rangle E_n^{(0)}(t)\langle n_0(t)|.
\eeq
The approximate time-dependent adiabatic solutions are
\beq
\label{a_s}
|\psi_n(t)\rangle=e^{i \xi_n(t)}|n_0(t)\rangle,
\eeq
where the adiabatic phase reads
\beq
\label{a_p}
\xi_n(t)=-\frac{1}{\hbar}\int_0^t dt' E_n^{(0)} (t')+ i \int_0^t dt' \langle n_0(t')|\partial_{t'} n_0(t')\rangle.
\eeq
Defining now the unitary operator
\beq
A(t)=\sum_n e^{i\xi_n(t)}|n_0(t)\rangle \langle n_0(0)|,
\eeq
a Hamiltonian $H(t)=i\hbar \dot A A^\dag$ can be constructed to drive the system exactly along the adiabatic paths of $H_0(t)$ as 
\beqa
H(t)&=&H_0(t)+H_{cd}(t),  \nonumber
\\
H_{cd}(t)&=&i\hbar \sum_n  ( |\dot{n}_0(t)\rangle \langle n_0(t)| \nonumber
\\
&-&\langle n_0(t)|\dot{n}_0(t)\rangle |n_0(t)\rangle \langle n_0(t)| ),
\eeqa
where $H_{cd}(t)$ is purely non-diagonal in the $\{ |n_0(t)\rangle\}$ basis and the dot represents time derivative.

For our system ($|n_0(t)\rangle\to|\phi_n\rangle$), the counterdiabatic term takes the form
\beq
H_{cd}=i\hbar (|\dot \phi_1\rangle \langle \phi_1|+|\dot \phi_3 \rangle \langle \phi_3|).
\eeq
Taking into account Eqs. (\ref{eigenstates_1}), (\ref{eigenstates_2}), (\ref{eigenstates_3}) and their respective time derivatives we get 
\beq
\label{H_cd}
H_{cd}=
-\hbar \dot \varphi G_2.
\eeq
Implementing this interaction is quite challenging as discussed in detail in \cite{Molmer}. 
In particular, a rapid switching between $G_1$ and $G_4$, to implement  $G_2$ through 
their commutator, is not a practical option \cite{Molmer}. Our goal in the following is to 
design an alternative Hamiltonan to perform the shortcut without $G_2$.     
\section{Alternative driving protocols via Lie transforms}
\label{as}
The main goal here is to define a new shortcut, different from the one described by $i\hbar \partial_t \psi(t)=H(t)\psi(t)$,
where $H(t)=H_0(t)+H_{cd}(t)$. 
A wave function  $\psi_I(t)$, that represents the alternative dynamics, is related to $\psi(t)$
by a unitary operator $B(t)$, 
\beq
\label{I_state}
\psi_I(t)=B^\dag(t)\psi(t), 
\eeq
and obeys 
$i\hbar \partial_t \psi_I(t)=H_I(t)\psi_I(t)$,
%
where   
\beqa
\label{I_hamiltonian}
H_I(t)&=&B^\dag(t)(H(t)-K(t))B(t),
\\
\label{I_K}
K(t)&=&i\hbar\dot B(t)B^\dag(t).
\eeqa
These are formally the same expressions that define  an interaction picture. However, in this application the 
``interaction picture'' represents a different physical setting from the original one \cite{IP}. 
In other words, $H_I$ is not a mathematical 
aid to facilitate a calculation in some transformed space, 
but rather a physically realizable Hamiltonian different from $H$. Similarly $\psi_I$ represents 
in general different dynamics from $\psi$.  
The transformation 
provides indeed an alternative shortcut   
if $B(0)=B(t_f)=1$, so that $\psi_I(t_f)=\psi(t_f)$
for a given initial state $\psi_I(0)=\psi(0)$. Moreover, if $\dot B(0)=\dot B(t_f)=0$ also the Hamiltonians coincide
at initial and final times, 
$H(0)=H_I(0)$ and $H(t_f)=H_I(t_f)$. These boundary conditions  may be relaxed 
in some cases as we shall see.   

We carry out the transformation by exponentiating a
member $G$ of the dynamical Lie algebra of the Hamiltonian, 
\beq\label{ag}
B(t)=e^{-i\alpha G},
\eeq
where $\alpha=\alpha(t)$ is a time dependent real function to be determined.  
This type of unitary operator $B(t)$ constitutes a ``Lie transform''. Lie transforms have been used for example  to 
develop efficient perturbative approaches that try to set the perturbation term of a
Hamiltonian in a convenient form both  in
classical and quantum systems \cite{classical,Bambusi}.
   
Note that $K$ in Eq. (\ref{I_K}) becomes $-\hbar \dot{\alpha}G$ and commutes with $G$.   
Then, $H_I$, given now by 
\beqa
B^\dag (H-K) B&=&e^{i\alpha G}(H-K)e^{-i\alpha G}
\nonumber
\\
&=&H-\hbar\dot{\alpha}G+i\alpha [G,H]-\frac{\alpha^2}{2!}[G,[G,H]]
\nonumber
\\
&-&i\frac{\alpha^3}{3!}[G,[G,[G,H]]] + \cdots
\label{tra}
\eeqa
depends only on $G$, $H$, and its repeated commutators with $G$, 
so it stays in the algebra. 
If we can choose $G$ and $\alpha$ so that the undesired generator components in $H$  cancel out 
and the boundary conditions 
for $B$ are satisfied, the method provides a feasible, alternative shortcut.  
In the existing applications of the method \cite{IP,review}, and in this paper
we proceed by trial an error, testing different generators. 
In the present application we want the Hamiltonian $H_I$ to keep the structure of the original one, with  
non-vanishing components proportional to $G_1$ and $G_4$.      
We may quickly discard by inspection
$G_1$, $G_2$, and $G_3$ as candidates for $G$.       
Choosing $G\to G_4$ in Eq. (\ref{ag}), 
and substituting into Eqs. (\ref{I_hamiltonian}) and (\ref{tra}),
the series of repeated commutators may be summed up.   
$H_I$ becomes  
\beqa
\label{H_I}
H_I&=&\left ( E_0 \cos \varphi-\hbar \dot \alpha \right ) G_4
\nonumber
\\
&-& \left ( E_0 \sin \varphi \cos \alpha + \hbar \dot \varphi \sin \alpha \right ) G_1
\nonumber
\\
&-& \left ( E_0 \sin \varphi \sin \alpha - \hbar \dot \varphi \cos \alpha \right ) G_2.
\eeqa
To cancel the $G_2$ term, we choose
\beq
\label{alpha}
\alpha(t)=\arccot \left [ \frac{E_0(t)}{\hbar \dot \varphi(t)} \sin(\varphi(t)) \right].
\eeq
Substituting Eq. (\ref{alpha}) into Eq. (\ref{H_I}) we have finally 
\begin{widetext}
\beq
\label{new_HI}
H_I=\left [ \frac{\cos \varphi E_0^3 \sin^2 \varphi +\hbar^2 \sin \varphi \dot E_0 \dot \varphi
 +\hbar^2 E_0 \left ( 2 \cos \varphi \dot \varphi ^2 - \sin \varphi \ddot \varphi \right )}{E_0^2 \sin^2 \varphi+ \hbar^2 \dot \varphi^2} \right ] G_4 
- \left [ E_0 \sin \varphi \sqrt{1+\frac{\hbar^2\csc^2\varphi \dot \varphi^2}{E_0^2}} \right ] G_1, 
\eeq
\end{widetext}
which has the same structure (generators) as the reference Hamiltonian but with different time-dependent coefficients. 
\section{Insulator-Superfluid transition}
\label{is}
Changing the $U/J$ ratio, the system may go 
from a ``Mott insulator'' (the two particles isolated in separate wells) to a ``superfluid'' state (in which each particle is distributed with equal probability in both wells).
From Eq. (\ref{eigenstates_1}),  the Mott-insulator ground state is  $|\phi_1\rangle=|1,1\rangle$ and in the superfluid regime the ground state becomes $|\phi_1\rangle=\frac{1}{2}|2,0\rangle+\frac{1}{\sqrt{2}}|1,1\rangle+\frac{1}{2}|0,2\rangle$.
To design a reference process (one that performs the transition when driven slowly enough)
we consider polynomial functions for $E_0(t)$ and $\varphi(t)$. 
Since we want to drive the system from $|1,1\rangle$ to $\frac{1}{2}|2,0\rangle+\frac{1}{\sqrt{2}}|1,1\rangle+\frac{1}{2}|0,2\rangle$, 
we impose in Eq. (\ref{eigenstates_1})
\beq
\label{bc_is}
\varphi(0)=0, 
\varphi(t_f) =\pi/2.
\eeq
To have the wells isolated at $t=0$ but connected (allowing the particles to pass from one to the other) at $t=t_f$ we also set
\beq
\label{bc_is2}
E_0(0)=0, 
E_0(t_f)\neq0,
\eeq
so that $J(0)=U(0)=0$ and $J(t_f)\neq0$.
Moreover, for a  smooth connection with the asymptotic regimes ($t<0$, $t>t_f$) we put  
\beq
\label{bc_is3}
\dot\varphi(0)=0, 
\dot\varphi(t_f) =0.
\eeq
This implies that $H_{cd}(0)=H_{cd}(t_f)=0$, see Eq. (\ref{H_cd}). 
The condition  
\beq
\label{bc_is4}
\ddot \varphi(t_f)=0
\eeq
is also needed to implement alternative shortcuts, in particular, to satisfy $\dot B(t_f)=0$.
At intermediate times, we interpolate the functions as  
$E_0(t)=\sum_{j=0}^1a_jt^j$ and $\varphi(t)=\sum_{j=0}^4b_jt^j$, where the coefficients are found by solving the equations for Eqs. (\ref{bc_is}), (\ref{bc_is2}), (\ref{bc_is3}) and (\ref{bc_is4}). These functions are shown in Fig. \ref{E0_phi_is}. In this and other figures 
$\tau=E_0^{max}t/\hbar$, where $E_0^{max}$ is the maximum value of $E_0(t)$. 
%
%
%
%
%
\begin{figure}[h]
\begin{center}
\includegraphics[height=2.9cm,angle=0]{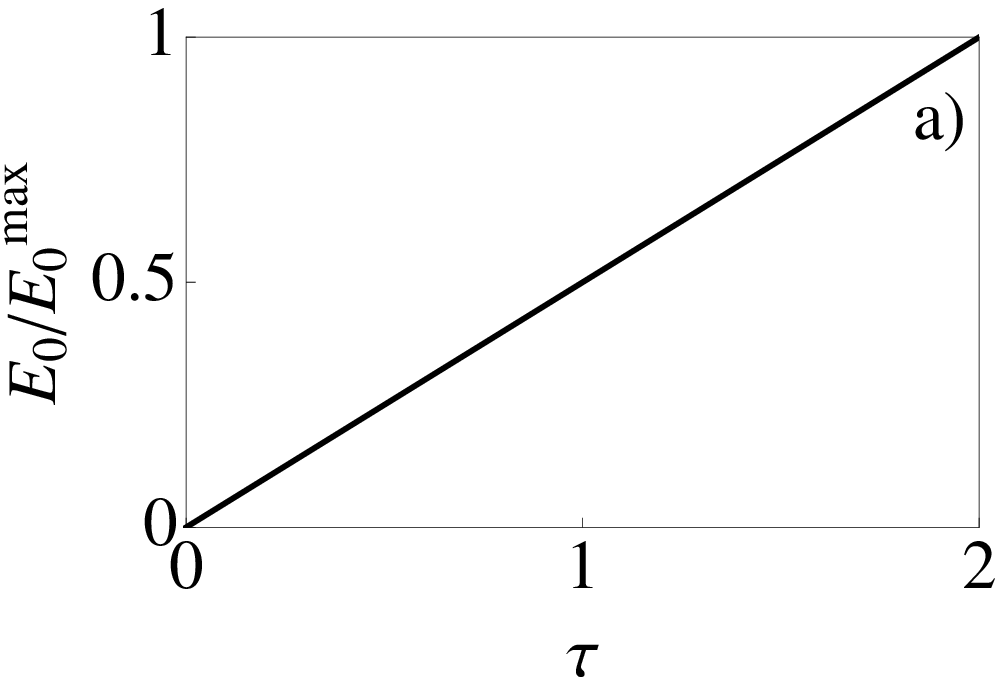}
\includegraphics[height=2.9cm,angle=0]{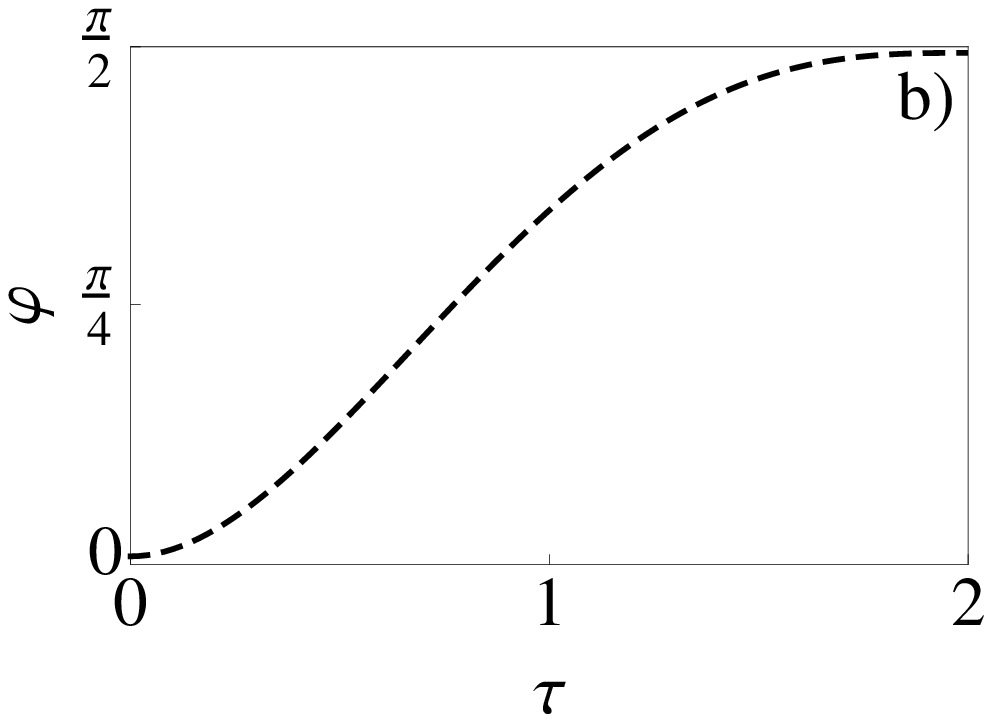}
\end{center}
\caption{\label{E0_phi_is}
Interaction Hamiltonian functions: (a) $E_0(t)$ and (b) $\varphi(t)$. Parameters: $\tau=E_0^{max}t/\hbar$ where $E_0^{max}$ is the maximum value of $E_0(t)$ and $\tau_f=2$.
 }
\end{figure}
%
%
%
%
%
%

The actual time evolution of the state
\beq
|\Psi(t)\rangle= c_1(t)|2,0\rangle + c_2(t)|1,1\rangle +c_3(t)|0,2\rangle,
\eeq
is given by solving Schr\"odinger's equation with the different Hamiltonians.  
For this particular transition, $|\Psi(0)\rangle=|\phi_1(0)\rangle$ and the ideal target state is
(up to a global phase factor) $|\Psi(t_f)\rangle=|\phi_1(t_f)\rangle$. 
%
%
%
%
\begin{figure}[h]
\begin{center}
\includegraphics[height=2.9cm,angle=0]{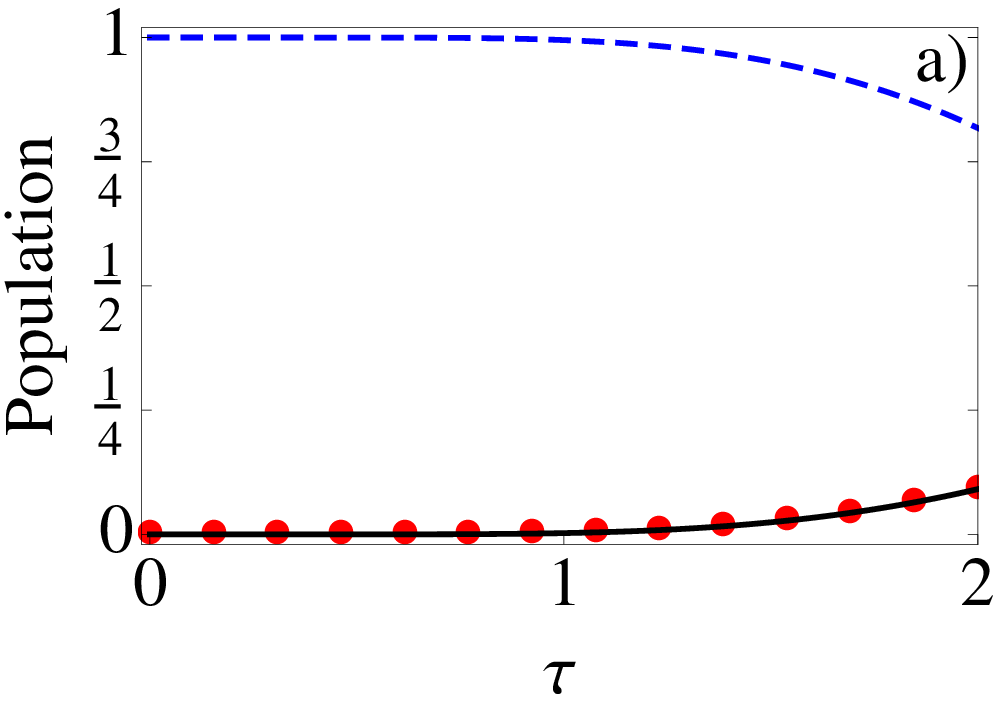}
\includegraphics[height=2.9cm,angle=0]{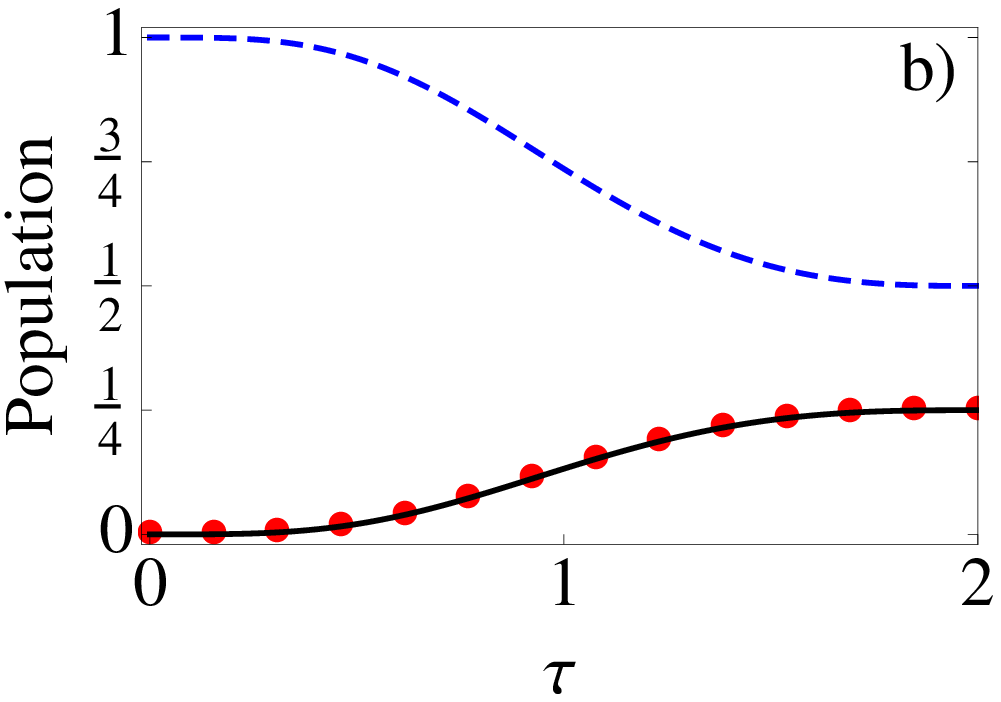}
\end{center}
\caption{\label{dynamics_is}
(Color online).
Bare-state populations for (a) $H_0(t)$, (b) $H(t)$ and $H_I(t)$. $|c_1(t)|^2$ (red circles), $|c_2(t)|^2$ 
(short-dashed blue line) and $|c_3(t)|^2$ (solid black line). Parameters: $\tau=E_0^{max}t/\hbar$ with $E_0^{max}$ the maximum value of $E_0(t)$, $\tau_f=2$. 
}
\end{figure}
%
%
%
%
%
%
The dynamics versus time $\tau$ is shown in Fig. \ref{dynamics_is} for $\tau_f=2$. 
For this short time $H_0(t)$ fails to drive the populations to $1/2$ and $1/4$, whereas when  
$H_{cd}(t)$ is added the intended transition occurs successfully.
As for the alternative Hamiltonian in Eq. (\ref{new_HI}),  
with $B=e^{-i\alpha G_4}$, and $\alpha$ in Eq. (\ref{alpha}),    
we find  
\beq
B(t_f)=1, 
\dot B(0)= \dot B(t_f)=0,
\eeq
(Eq. (\ref{bc_is4}) is necessary to have $\dot \alpha (t_f)=0$ and consequently $\dot B(t_f)=0$), 
whereas 
\beq
\label{B_is}
B(0)=\left ( \begin{array}{ccc}
e^{-i\pi/2} & 0 & 0 \\
0 & 1 & 0 \\
0 & 0 & e^{-i\pi/2} \end{array} \right )\ne 1.
\eeq
However $B^\dagger(0)|1,1\rangle=|1,1\rangle$ so $\psi^I(0)=\psi(0)$ and $H_I$ provides the desired shortcut.

Solving numerically the dynamics for $H_I(t)$ we obtain a perfect insulator-superfluid transition (see Fig. \ref{dynamics_is} (b)). Notice that, as $G_4$ is diagonal in the bare basis, the bare-populations are the same for the dynamics driven by $H$ and $H_I$, see Fig. \ref{dynamics_is} (b).
%
%
%
%
%
%
%
%
%
\begin{figure}[h]
\begin{center}
\includegraphics[height=2.9cm,angle=0]{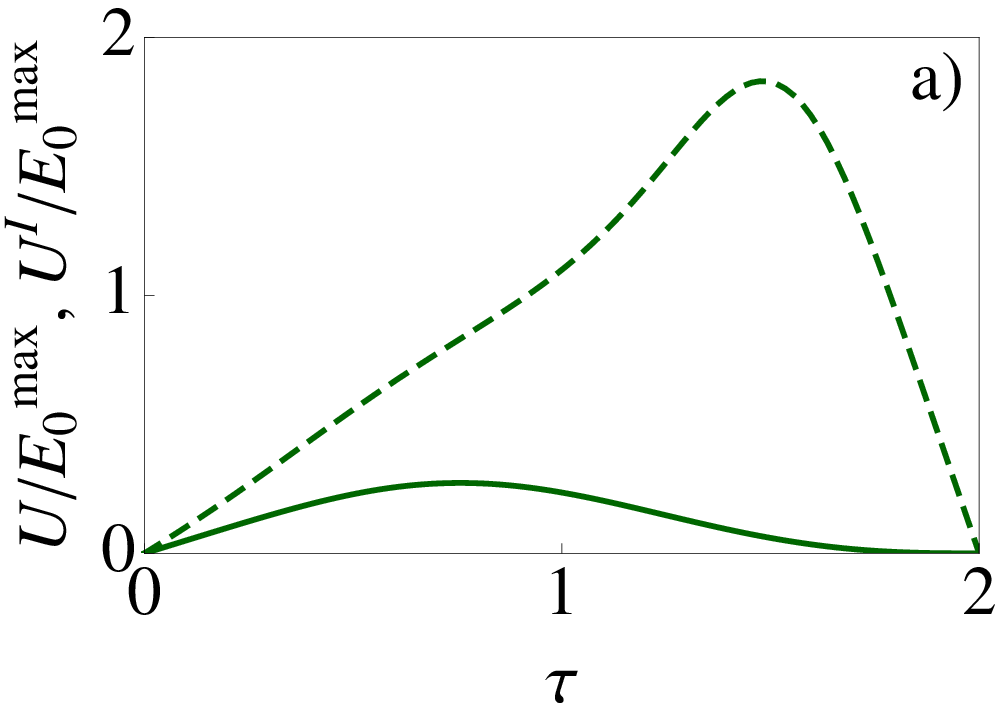}
\includegraphics[height=2.9cm,angle=0]{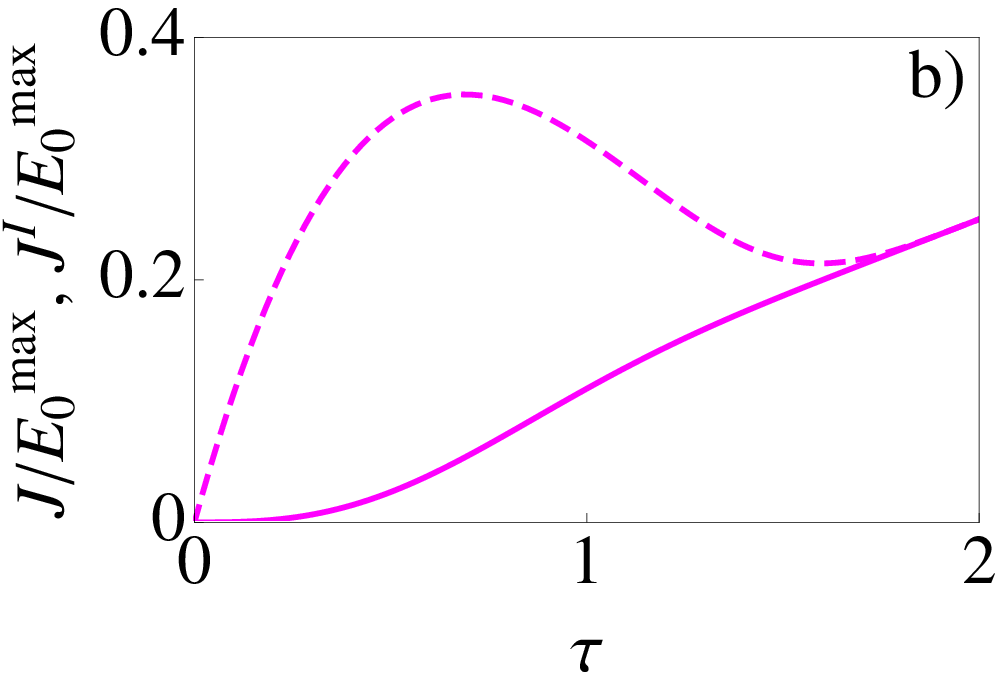}
\end{center}
\caption{\label{U_J_is}
(Color online).
(a) Interaction energy for the reference (solid green line) and interaction (short-dashed green line) Hamiltonian. (b) Hopping energy for the reference (solid magenta line) and interaction (short-dashed magenta line) Hamiltonian. Same parameters as Fig. \ref{E0_phi_is}.
 }
\end{figure}
%
%
%
%
%
%

In order to compare our approach 
with other protocols 
we reformulate $H_I$ as
\beq
\label{H_Ir}
H_I=\left ( \begin{array}{ccc}
U^I & -\sqrt{2}J^I & 0\\
-\sqrt{2}J^I & 0 & -\sqrt{2}J^I\\
0 & -\sqrt{2}J^I & U^I
\end{array} \right)=U^I G_4 - 4 J^I G_1.
\eeq
Comparing Eqs. (\ref{H_Ir}) and (\ref{new_HI}) we find that
\beqa
U^I&=&\frac{1} {{(E_0)}^2 \sin^2 \varphi+ \hbar^2 {(\dot \varphi)}^2} 
\left [ \cos \varphi{(E_0)}^3 \sin^2 \varphi\right.
\nonumber\\
&+&\left.\hbar^2 \sin \varphi \dot E_0 \dot \varphi+\hbar^2 E_0 \left ( 2 \cos \varphi{(\dot \varphi)}^2 - \sin \varphi \ddot \varphi\right ) \right ] , 
\nonumber\\ 
J^I&=&\frac{1}{4} E_0\sin \varphi \sqrt{1+\frac{\hbar^2\csc^2\varphi {(\dot \varphi)}^2}{{(E_0)}^2}}.
\eeqa
Figure \ref{U_J_is} shows the functions $U_I$ and $J_I$. We have set $H_I(t_b)=H_0(t_b)$, for $t_b=0,t_f$, since  $H_{cd}(t_b)=0$ and 
$\dot{B}(t_b)=0$. 
In the same way as Eq. (\ref{parametrization_U}) 
we can rewrite the above energies as
\beq
\label{new_parametrizationU}
U^I=E_0^I \cos\varphi^I, \;
J^I=\frac{E_0^I}{4} \sin \varphi^I,
\eeq
where $E_0^I=E_0^I(t)$ and $\varphi^I=\varphi^I(t)$.  
The inverse transformation is 
\beq
\label{new_parametrizationphi}
\varphi^I=\arctan\left ({4\frac{J^I}{U^I}}\right ), \;
E_0^I=\frac{U^I}{\cos \varphi'}.
\eeq
Consider a simple  protocol with $E_0(t)=E_0^M(t)=const.$ and a linear $\varphi^M(t)$ from  $0$ and $\pi/2$ \cite{Molmer}. 
Setting the value of $E_0^M$ so that   $\int E_0^M dt=\int E^I_0dt$, it is found that the simple protocol
needs $\tau_f=18.8$ to perform the transition with a 0.9999 fidelity.
In other words, the protocol based on $H_I$ is $9.4$ times faster according to this criterion. 
\section{Beam splitters}
\label{bs}
The three-level Hamiltonian (\ref{H_0}) describes other physical systems apart from two bosons in two wells. 
For example it represents in the paraxial approximation and substituting time by a longitudinal coordinate
three coupled waveguides
\cite{Longhi_2008,Rev_Longhi,Rev_Nolte,Longhi_2011,Vitanov_2012, Tseng_2013}, 
where $J$ is controlled by waveguide separation and $U$ by the 
refractive index. In particular $J$ and $U$ may be manipulated to split an incoming wave in the central wave guide into two output channels
 (corresponding to the external waveguides) or 
three output chanels   \cite{Vitanov_2012, Tseng_2013}.
The Hamiltonian also  
represents a single particle in a triple well \cite{Mompart_2004},
with $U$ representing the bias of the outer wells with respect to the central one and $J$
the coupling coefficient between adjacent wells. The beam splitting may thus represent the evolution 
of the particle wave function from the central well either to the two outer wells 
or to  three of them with equal probabilities.

For either of these physical systems\footnote{The Hamiltonian (\ref{H_0}) also describes a three-level atom
under appropriate laser interactions, see \cite{Longhi_2008}.} 
the minimal channel basis for left, center and right wave functions is $|L\rangle=\left ( \scriptsize{\begin{array} {rcccl} 1\\ 0\\0 \end{array}} \right)$, $|C\rangle=\left ( \scriptsize{\begin{array} {rcccl} 0\\ 1\\0 \end{array}} \right)$ and 
$|R\rangle=\left ( \scriptsize{\begin{array} {rcccl} 0\\ 0\\1 \end{array}} \right)$. 
\subsection{1:2 beam splitter}
%
%
%
%
%
%
\begin{figure}[h]
\begin{center}
\includegraphics[height=2.4cm,angle=0]{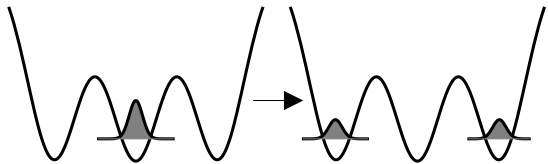}
\end{center}
\caption{\label{lr}
Schematic representation of a $1:2$ beam splitter.}
\end{figure}
%
%
%
%
%
%
%
To implement a $1:2$ beam splitter, see Fig. \ref{lr},  
the goal is to drive the eigenstate from $|\phi_1(0)\rangle=|C\rangle$ to $|\phi_1(t_f)=\frac{1}{\sqrt{2}}\left ( |L\rangle+|R\rangle \right )$. 
%
%
As in the previous section we use polynomial functions for $E_0(t)$ and $\varphi(t)$ to set a reference process.
We impose
\beq
\label{bc_lr}
\varphi(0)=0, \varphi(t_f) =\pi
\eeq
in Eq. (\ref{eigenstates_1}). 
The wells (waveguides) should be isolated at initial and final times. 
If morever all wells are at equal heights at those times we set
\beq
\label{bc_lr2}
E_0(0)=E_0(t_f) =0, E(t_f/2)\neq0,
\eeq
to satisfy $H_0(0)=H_0(t_f)=0$.
We  also impose
\beq
\label{bc_lr3}
\dot\varphi(0)=0, \dot\varphi(t_f) =\pi
\eeq
to smooth the functions at the time boundaries and make $H_{cd}(t_b)=0$. 
In addition  
\beq
\label{bc_lr4}
\ddot \varphi(t_f)=0
\eeq
is imposed to satisfy $\dot B(t_f)=0$. 
At intermediate times  $E_0(t)=\sum_{j=0}^2a_jt^j$ and $\varphi(t)=\sum_{j=0}^4b_jt^j$, with the coefficients 
deduced from  Eqs. (\ref{bc_lr}), (\ref{bc_lr2}), (\ref{bc_lr3}) and (\ref{bc_lr4}).
These functions are shown in Fig. \ref{E0_phi_LR}.
%
%
%
%
%
\begin{figure}[h]
\begin{center}
\includegraphics[height=2.9cm,angle=0]{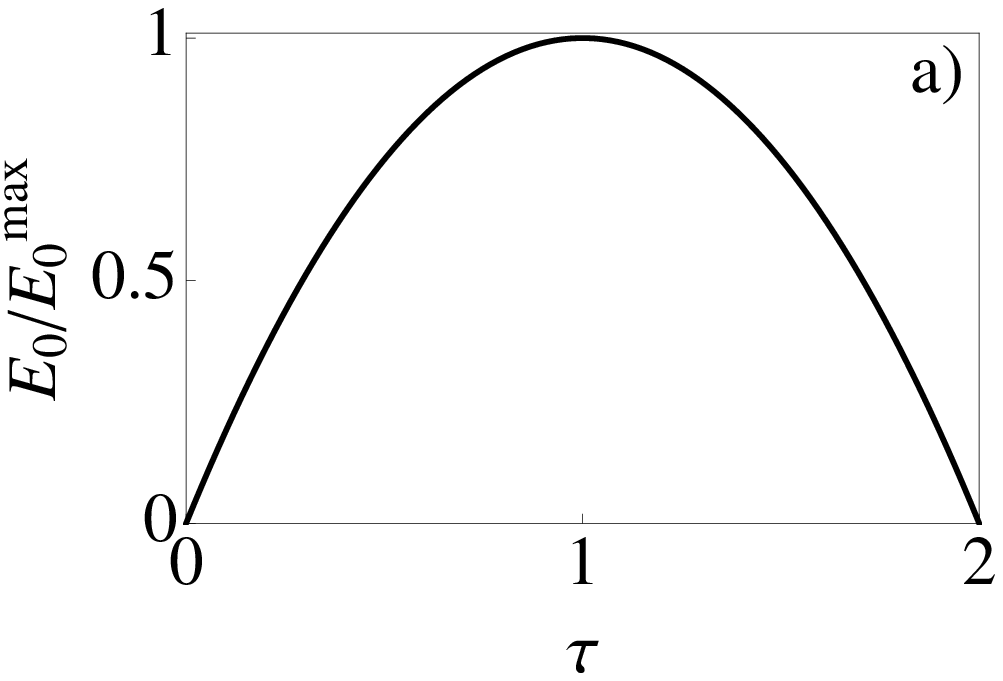}
\includegraphics[height=2.9cm,angle=0]{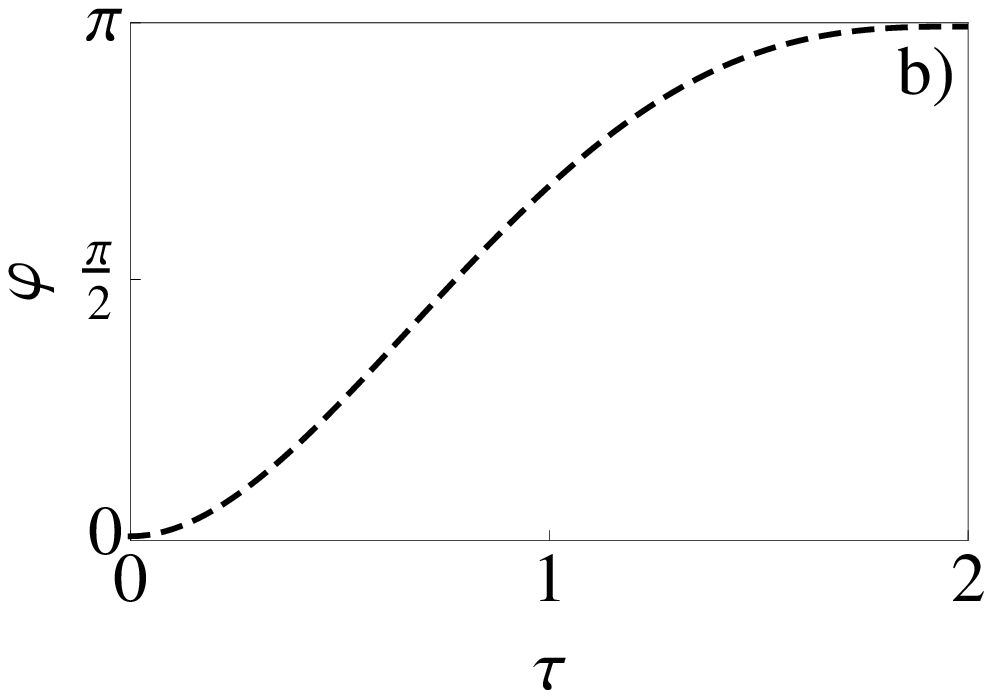}
\end{center}
\caption{\label{E0_phi_LR}
(a) $E_0(t)$ and (b) $\varphi(t)$. $\tau=E_0^{max}t/\hbar$ where $E_0^{max}$ is the maximum value of $E_0(t)$. $\tau_f=2$.
 }
\end{figure}
%
%
%
%
%
%
%
%
%
%
%
%
\begin{figure}[h]
\begin{center}
\includegraphics[height=2.9cm,angle=0]{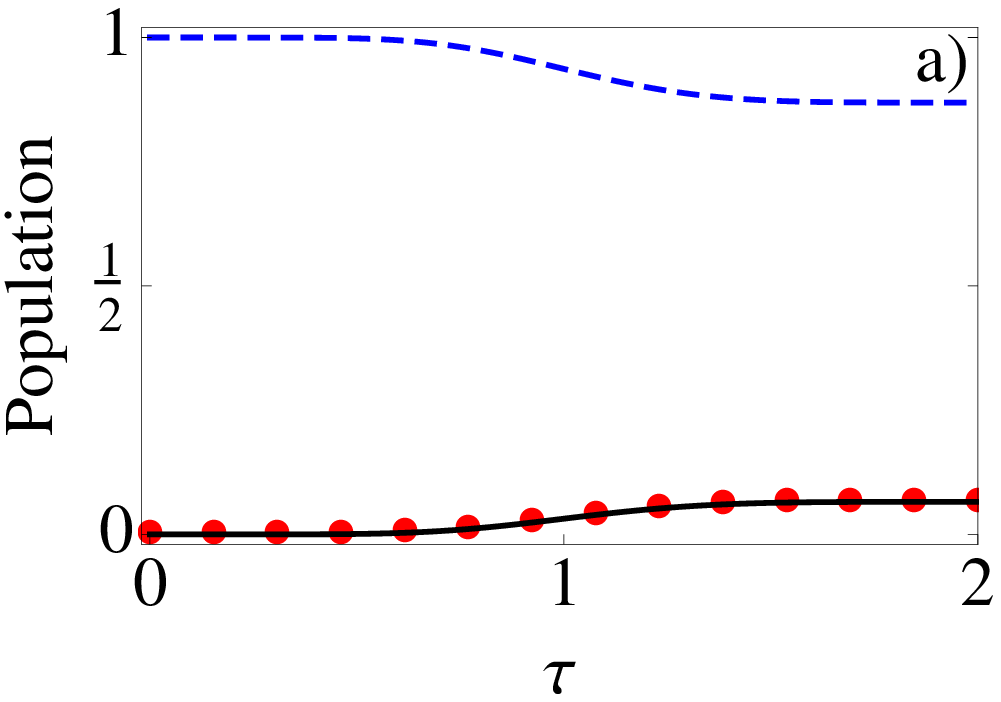}
\includegraphics[height=2.9cm,angle=0]{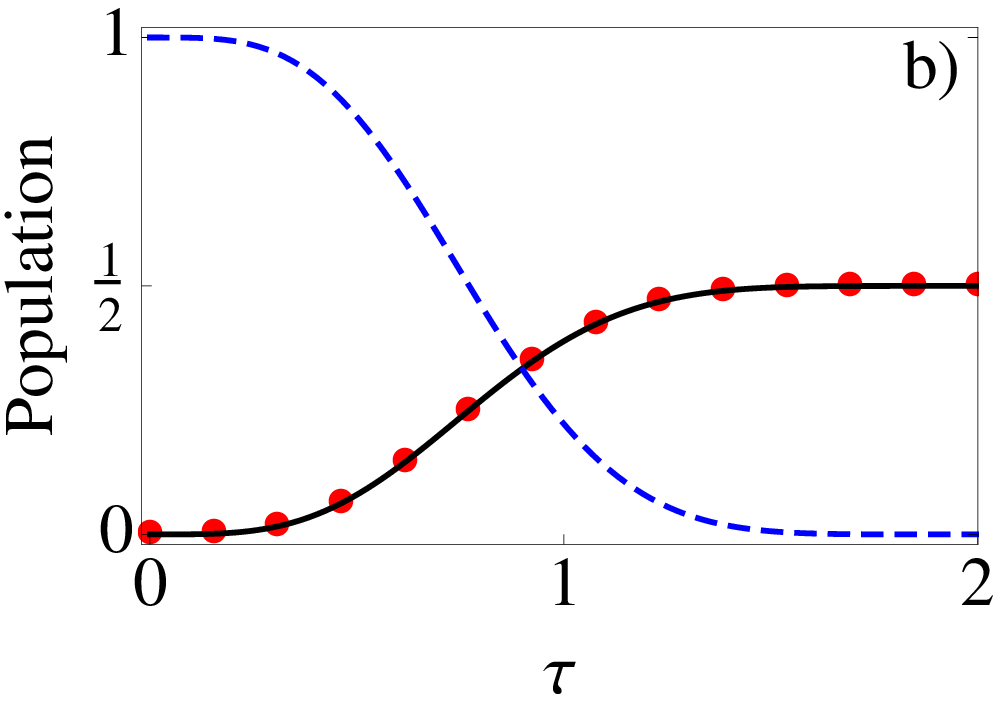}
\end{center}
\caption{\label{dynamics_LR}
(Color online).
Bare-state populations for (a) $H_0(t)$, (b) $H(t)$ and $H_I(t)$. $|c_1(t)|^2$ (red circles), $|c_2(t)|^2$ 
(short-dashed blue line) and $|c_3(t)|^2$ (solid black line). Parameters: $\tau=E_0^{max}t/\hbar$ with $E_0^{max}$ the maximum value of $E_0(t)$, $\tau_f=2$}
\end{figure}
%
%
%
%
%
%

Figure \ref{dynamics_LR} shows the dynamics  for $\tau_f=2$. This time (corresponding to the splitter length in the optical system) is too short for the reference Hamiltonian $H_0(t)$ to drive the bare-basis populations to $0$ and $1/2$. Adding $H_{cd}(t)$ the transition occurs as desired. As in Sec. \ref{as}, we  construct an alternative shortcut $H_I(t)$ without $G_2$ 
using the transformation $B=e^{-i\alpha G_4}$. With $\alpha$ in Eq. (\ref{alpha}),  
$\dot B(0)=\dot B(t_f)=0$, whereas
%
\beq
\label{B_lr}
B(0)=B(t_f)=\left ( \begin{array}{ccc}
e^{-i\pi/2} & 0 & 0 \\
0 & 1 & 0 \\
0 & 0 & e^{-i\pi/2} \end{array} \right ).
\eeq
This is enough for our objective as $B^\dagger(0)|C\rangle=|C\rangle$, and $B^\dag(t_f) |\psi(t_f)\rangle=-i|\psi(t_f)\rangle$. 
%
%
%
%
%
\begin{figure}[h]
\begin{center}
\includegraphics[height=2.8cm,angle=0]{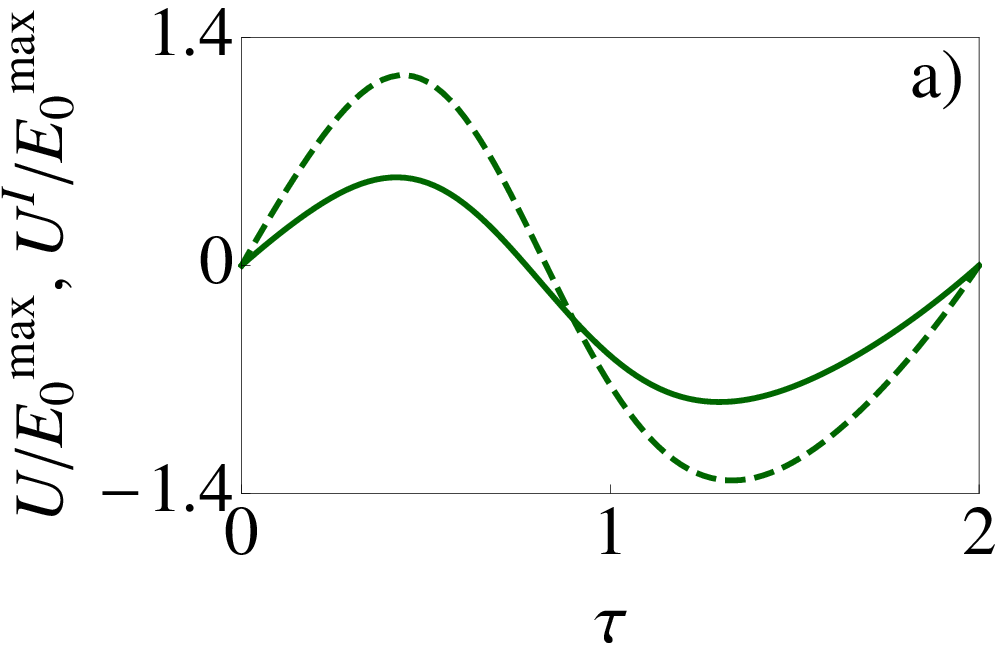}
\includegraphics[height=2.8cm,angle=0]{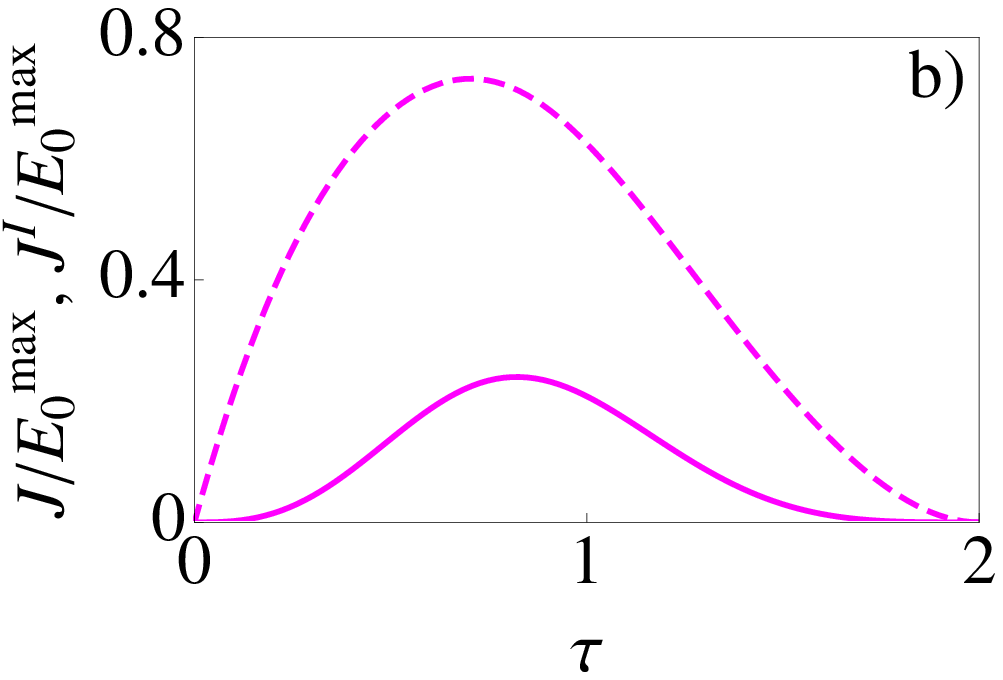}
\end{center}
\caption{\label{U_J_LR}
(Color online).
(a) Interaction energy for the reference (solid green line) and interaction (short-dashed green line) Hamiltonian. (b) Hopping energy for the reference (solid magenta line) and interaction (short-dashed magenta line) Hamiltonian. Same parameters as Fig. \ref{E0_phi_LR}
 }
\end{figure}
%
%
%
%
%
%
Solving numerically the dynamics for $H_I(t)$ we obtain a perfect $1:2$ beam splitting (see Figs. \ref{U_J_LR}  and \ref{dynamics_LR} (c)).

To compare the new shortcut and the simple approach with $E_0^M=const.$ and $\varphi^M(t)=\frac{t}{t_f}\pi$ we 
set  
$\int E_0^M dt=\int E_0^I dt$. The constant-$E_0$ protocol needs $\tau_f \geq 18.6$ to achieve $0.9999$ fidelity, 
so the protocol driven by $H_I$ is $9.3$ times faster. 
\subsection{1:3 beam splitter}
%
%
%
%
%
%
\begin{figure}[h]
\begin{center}
\includegraphics[height=2.4cm,angle=0]{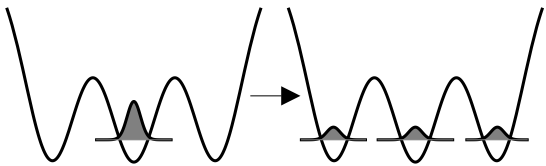}
\end{center}
\caption{\label{lcr}
Schematic representation of the beam splitter $1:3$.}
\end{figure}
%
%
%
%
%
%
%
%
We also describe briefly a $1:3$ beam splitter, see Fig. \ref{lcr}.   
The aim is to drive the system from $|\phi_1(0)\rangle=|C\rangle$ to equal populations in 
$|L\rangle$, $|C\rangle$, and $|R\rangle$. To design a reference protocol we use polynomial interpolation for $E_0(t)$ and $\varphi(t)$, see Fig. \ref{E0_phi_LCR}, 
with the same boundary conditions of the $1:2$ splitter but with $\varphi(t_f) =0.60817\pi=\arccos(-1/3)$ and the additional condition $\dot E_0(t_f)=0$ (to satisfy $U^I(t_f)=U(t_f)$ so that $H_I(t_f)=H_0(t_f)$).  The Lie  transform may be applied as before on the protocol with the counterdiabatic correction, see  Fig. \ref{dynamics_LCR} (b). 
%
%
%
%
%
%
\begin{figure}[h]
\begin{center}
\includegraphics[height=2.9cm,angle=0]{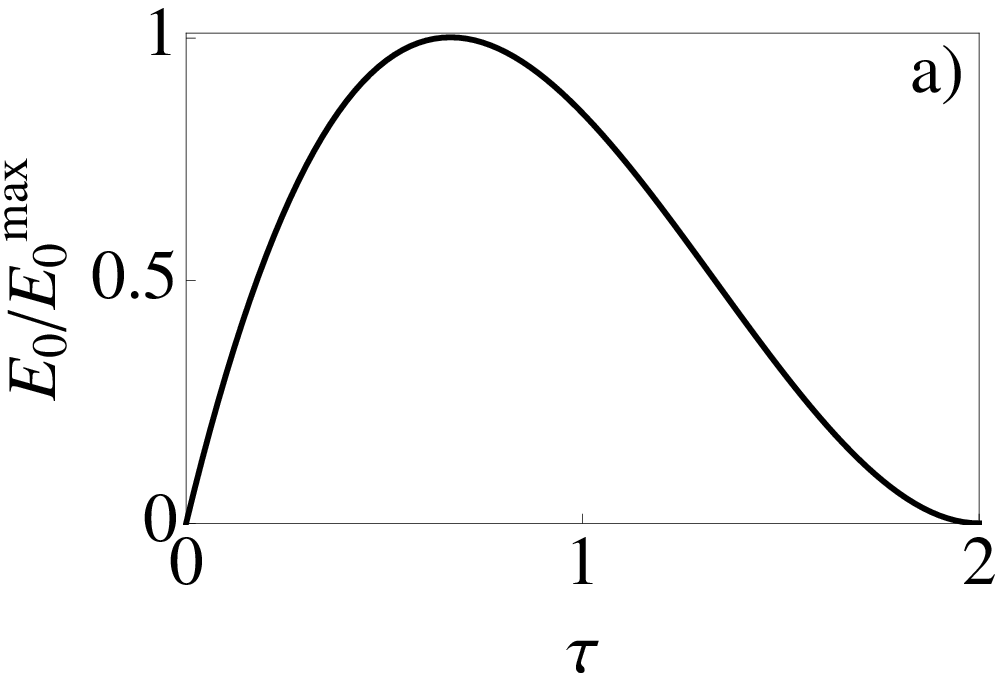}
\includegraphics[height=2.9cm,angle=0]{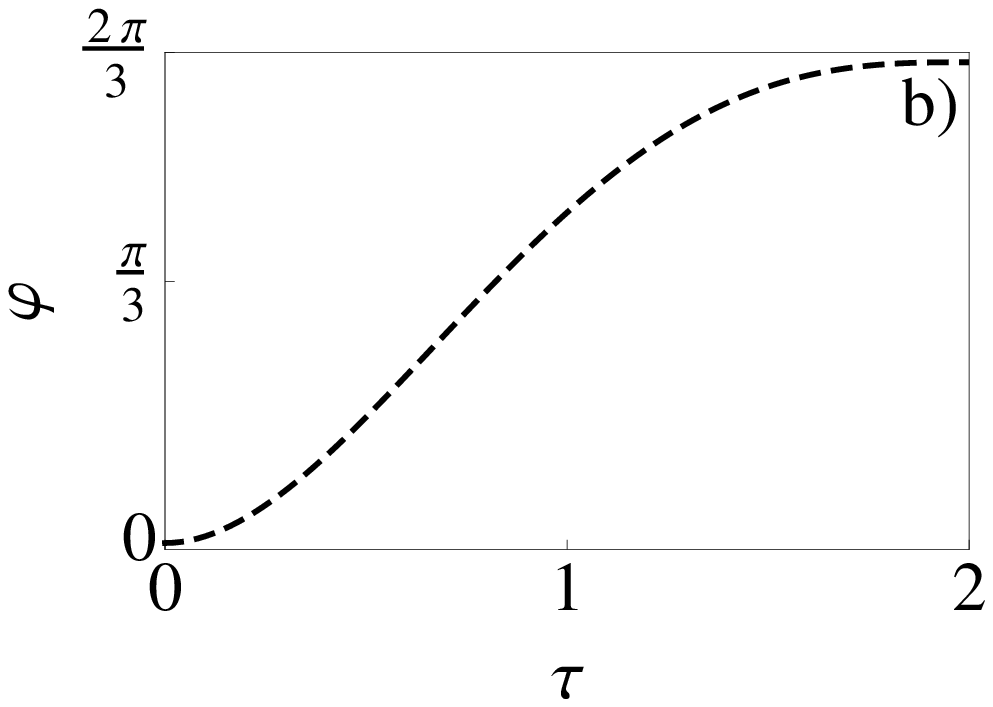}
\end{center}
\caption{\label{E0_phi_LCR}
(a) $E_0(t)$ and (b) $\varphi(t)$. $\tau=E_0^{max}t/\hbar$, where $E_0^{max}$ is the maximum value of $E_0(t)$. $\tau_f=2$.
 }
\end{figure}
%
%
%
%
%
%
A simple protocol
with $E_0^M$ and $\varphi(t)=\frac{t}{t_f} 0.60817\pi$
needs  $\tau_f=22$, if  $\int E_0^M dt=\int E_0^I dt$, for a 0.9999 fidelity, 
so the protocol based on $H_I$ is 11 times faster. 
%
%
%
%
%
%
%
\begin{figure}[h]
\begin{center}
\includegraphics[height=2.9cm,angle=0]{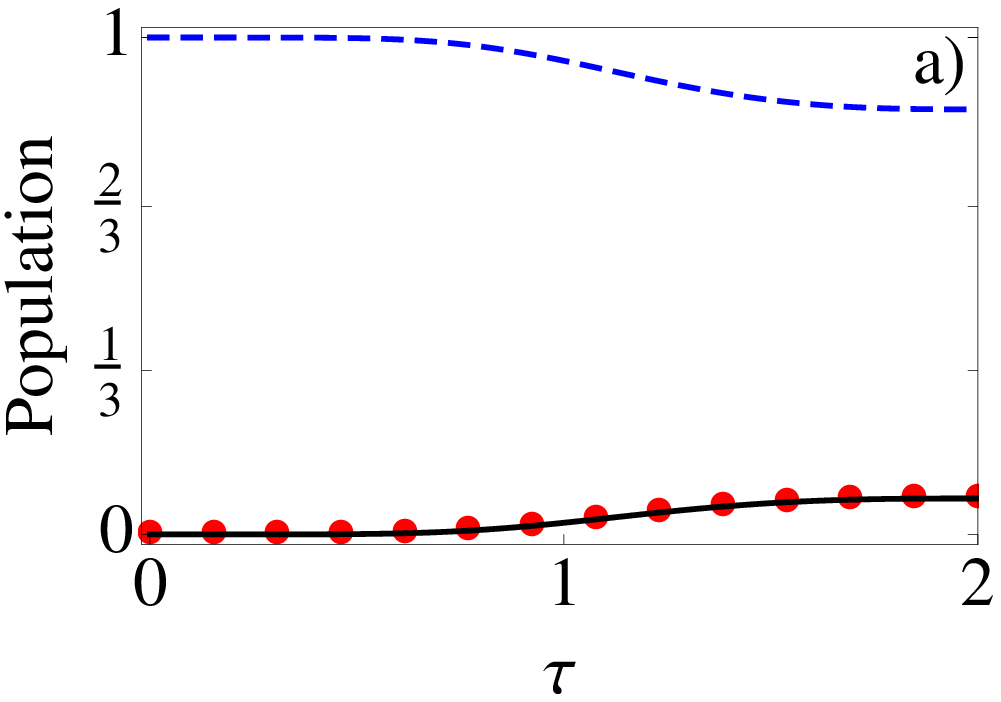}
\includegraphics[height=2.9cm,angle=0]{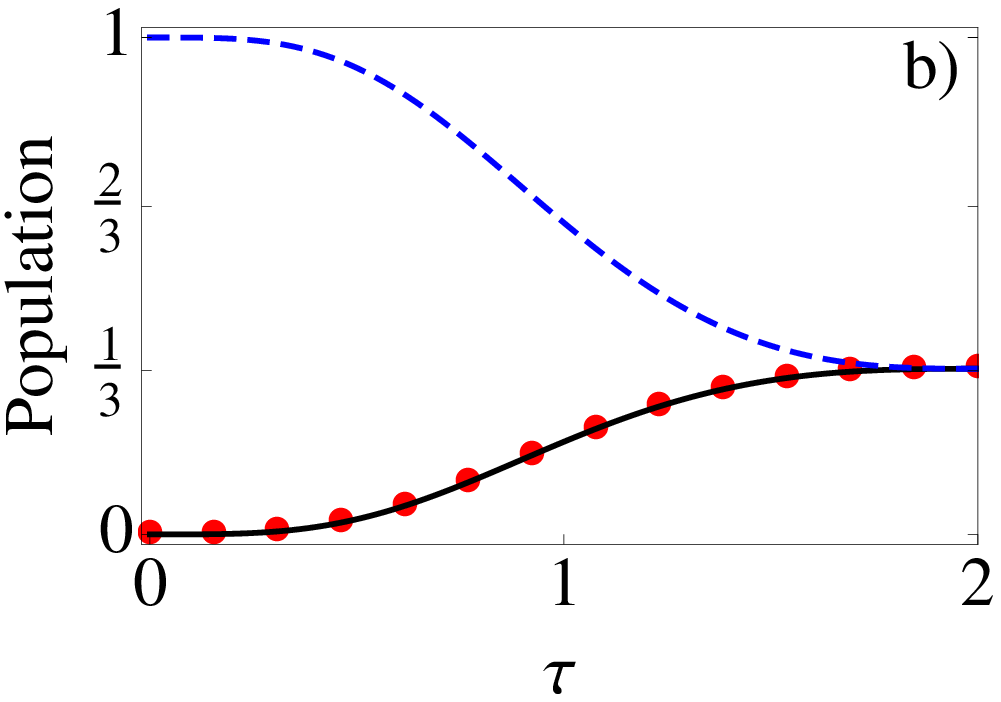}
\end{center}
\caption{\label{dynamics_LCR}
(Color online).
Bare-state populations for (a) $H_0(t)$, (b) $H(t)$ and $H_I(t)$. $|c_1(t)|^2$ (red circles), $|c_2(t)|^2$ 
(short-dashed blue line) and $|c_3(t)|^2$ (solid black line). Parameters: $\tau=E_0^{max}t/\hbar$ with $E_0^{max}$ the maximum value of $E_0(t)$, $\tau_f=2$.}
\end{figure}
%
%
%
%
%
%
%
%
%
%
\begin{figure}[h]
\begin{center}
\includegraphics[height=2.8cm,angle=0]{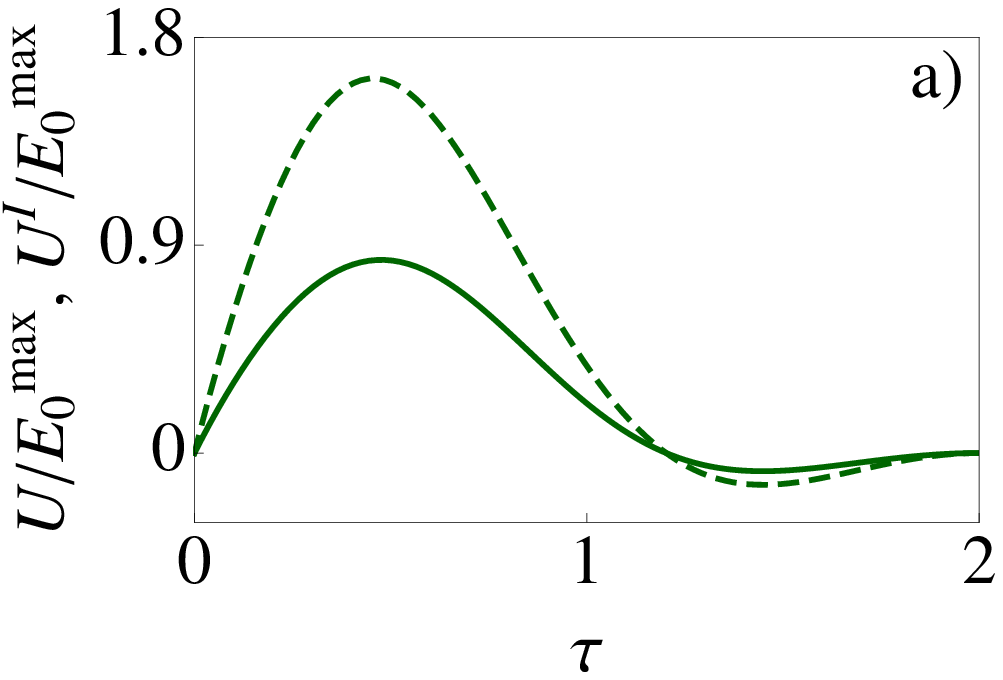}
\includegraphics[height=2.8cm,angle=0]{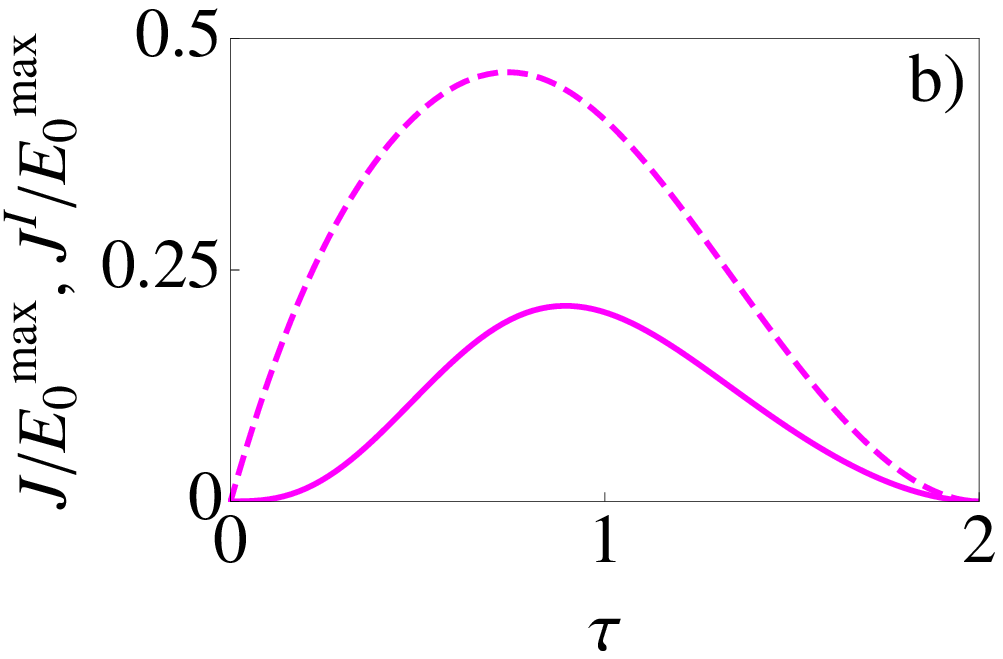}
\end{center}
\caption{\label{U_J_LCR}
(Color online).
(a) Interaction energy for the reference (solid green line) and interaction (short-dashed green line) Hamiltonian. (b) Hopping energy for the reference (solid magenta line) and interaction (short-dashed magenta line) Hamiltonian. Same parameters as in Fig. \ref{E0_phi_LCR}
 }
\end{figure}

%
%
%
%
%
%

\section{Discussion}
\label{discussion}
Starting from shortcuts to adiabaticity for three-level systems with U3S3 symmetry 
(a four-dimensional Lie algebra) 
that include Hamiltonian terms difficult to implement in the laboratory, 
we have found alternative shortcuts without them by means of Lie transforms.  
These transformations are formally equivalent to interaction picture (IP) transformations.
However the resulting IP-Hamiltonian and state represent a different
physical process from the original  (Schr\"odinger) Hamiltonian and dynamics.      
We have found shortcuts for different physical systems. For two particles in two wells we have implemented 
a fast insulator-superfluid transition. For coupled waveguides or a particle in a triple well we have implemented 
fast beam splitting with one input channel and two or three output channels. In all cases the IP Hamiltonian  
involves only two realizable terms (generators).  

In a companion paper we have worked out a related method  \cite{Lie}. 
Both approaches rely on Lie algebraic methods 
and aim at constructing shortcuts to adiabaticity. However we do not use dynamical invariants
explicitly in the current approach,  whereas the bottom-up approach in 
\cite{Lie} engineers the Hamiltonian making explicit use of     
its relation to dynamical invariants. In contrast we start here from an existing, known shortcut
--for example the one generated by a counter-diabatic method--;  
then, a Lie transform is applied to generate alternative, 
feasible or more convenient shortcuts, as in \cite{IP}.     
A connection between the transformation method and dynamical invariants is sketched briefly in the
Appendix but it deserves an extensive separate study.  

Finally, further applications of this work may involve systems with Lie algebras of higher dimension. 
Within the scope of the  algebra U3S3, other physical systems that could be treated are 
in quantum optics (three level atoms) 
 \cite{Chen_3ls, rev_Shore}, nanostructures (triple wells or dots) \cite{Kiselev}, optics 
(mode converters) \cite{Tseng_2012,Chen_2012}, or Bose-Einstein condensates in an accelerated optical lattice
\cite{Dou}. 

\section*{Acknowledgments}
This work was supported by the National Natural Science Foundation
of China (Grant No. 61176118), 
the Grants
No. 12QH1400800,
IT472-10, BFI-2010-255, 13PJ1403000,
FIS2012-36673-C03-01, and the program UFI 11/55.
S. M.-G. acknowledges a fellowship by UPV/EHU.  E. T. is supported by the Basque Government
postdoctoral program. 
\appendix
\section{Lie algebra}
\label{algebra}
The algebra of this three-level system is a four-dimensional Lie Algebra U3S3 according to the classification of 4-dimensional Lie algebras in \cite{Mac}. 
(For comparison with that work it is useful to rewrite the generators in the skew-Hermitian base  $\tilde G_k=-i G_k$, $k=1,2,3,4$.) 
$U3S3$ 
is a direct sum of the one dimensional algebra spanned by the invariant $G_4-G_3$, that commutes with all members of the algebra, 
and a three-dimensional SU(2) algebra spanned by $\{G_1, G_2, G_3\}$.  
Notice that this realization of the 3D algebra is not spanned by  the matrices 
\beqa
\label{generators_spin1}
J_x&=&\frac{1}{\sqrt{2}}\left ( \begin{array}{ccc}
0 & 1 & 0\\
1 & 0 & 1 \\
0 & 1 & 0
\end{array} \right), \nonumber
\\ \nonumber
\\
J_y&=&\frac{1}{\sqrt{2}}\left ( \begin{array}{ccc}
0 & -i & 0\\
i & 0 & -i \\
0 & i & 0
\end{array} \right), 
J_z=\left ( \begin{array}{ccc}
1 & 0 & 0\\
0 & 0 & 0 \\
0 & 0 & -1
\end{array} \right), 
\eeqa
that correspond, in the subspace ${|2,0\rangle,|1,1\rangle,|0,2\rangle}$, to the operators
\beqa
\label{spin1}
J_x&=&\frac{1}{2}\left ( a_1^\dag a_2+a_2^\dag a_1 \right ), 
\\
J_y&=&\frac{1}{2i}\left ( a_1^\dag a_2-a_2^\dag a_1 \right ), 
\\
J_z&=&\frac{1}{2} \left ( a_1^\dag a_1-a_2^\dag a_2 \right ).
\eeqa
In particular we cannot get the matrices for $J_y$ or $J_z$ by any linear combination of our $G_k$
matrices (see Eqs. (\ref{G1_G4}-\ref{generators})).
A second-quantized form for the $G_k$ consistent with the matrices includes quartic terms in annihilation/creation operators: 
%
%
\beqa
G_1&=&\frac{1}{4}\left ( a_1^\dag a_2+a_2^\dag a_1 \right ), \nonumber
\\
G_2&=&\frac{1}{4i}\left [ a_1^\dag a_2^\dag  \left ( a_1 a_1 + a_2 a_2 \right )\right. 
\nonumber\\
&-& \left.\left (a_1^\dag a_1^\dag+ a_2^\dag a_2^\dag \right ) a_1 a_2  \right ], \nonumber
\\
G_3&=&\frac{1}{8} \left [ \left (a_1^\dag a_1^\dag+ a_2^\dag a_2^\dag \right) a_1 a_1 - 4 a_1^\dag a_2^\dag a_1 a_2\right. 
\nonumber\\
&+&  \left.\left (a_1^\dag a_1^\dag+ a_2^\dag a_2^\dag \right ) a_2 a_2 \right ] , \nonumber
\\
G_4&=&\frac{1}{4} {\left( a_1^\dag a_1 - a_2^\dag a_2 \right )} ^2, 
\eeqa
%
Notice that these second-quantized operators do not form a closed algebra under commutation but their matrix elements
for two particles do. 

An invariant (defined in a Lie-algebraic sense) commutes with any member  of the algebra.
There are generically two independent invariants for $U3S3$ \cite{patera}.   
For the matrix representation in Eqs. (\ref{G1_G4}) and (\ref{generators}) they are 
\beqa
I_1&=&G_1^2=G_2^2=G_3^2=\frac{1}{8}\left ( \begin{array}{ccc}
1 & 0 & 1\\
0 & 2 & 0 \\
1 & 0 & 1
\end{array} \right),
\nonumber
\\
I_2&=&G_4-G_3=\frac{1}{4}\left ( \begin{array}{ccc}
3 & 0 & -1\\
0 & 2 & 0\\
-1 & 0 & 3
\end{array} \right).
\eeqa
$I_1$, which is not in the algebra,  has eigenvalues
\beq
\lambda_1^{(2)}=1,\; 
\lambda_1^{(1,3)}=\frac{1}{2}, 
\eeq
and $I_2$, a member of the algebra,  has eigenvalues
\beq
\lambda_2^{(2)}=0,
\;
\lambda_2^{(1,3)}=\frac{1}{4}. 
\eeq
The two invariants have the same eigenvectors,
\beqa
|u^{(1)}\rangle&=&\frac{1}{\sqrt{2}}(|2,0\rangle + |0,2\rangle),
\nonumber
\\
|u^{(2)}\rangle&=&\frac{1}{\sqrt{2}}(|2,0\rangle- |0,2\rangle),
\nonumber
\\
|u^{(3)}\rangle&=&|1,1\rangle.
\eeqa
with $|u^{(1)}\rangle$ and $|u^{(3)}\rangle$ spanning a degenerate subspace.

Lie-algebraic invariants constructed with time-independent coefficients 
satisfy  as well the equation 
\beq
\label{di}
i\hbar \frac{\partial I_{1,2}}{\partial t} +  [H(t),I_{1,2}]=0
\eeq
so they are also dynamical invariants \cite{LR} (i.e., operators that satisfy  Eq. (\ref{di}) whose expectation values remain constant). 
The  degenerate subspace of eigenvectors allows the existence of time-dependent eigenstates 
of time-independent invariants. 
In particular, in all the examples in the main text, the dynamics takes place within the degenerate subspace:  
the initial state is $|u^{(3)}\ra$ at $t=0$ and ends up in some combination of $|u^{(1)}\ra$ and $|u^{(3)}\ra$ at $t_f$.
The specific state as a function of time is known explicitly, 
$|\psi_I(t)\ra=e^{i\alpha(t)G_4}e^{-i\int_0^{t} E_1 dt'} |\phi_1(t)\ra$, see Eq. (\ref{I_state}).   
Note that $|\phi_1\ra$ and $|\phi_3\ra$ in Eqs. (\ref{eigenstates_1}) and (\ref{eigenstates_3}) 
are two orthogonal combinations of $|u^{(1)}\ra$ and $|u^{(3)}\ra$. Also 
$|u^{(2)}\ra=|\phi_2\ra$, see Eq. (\ref{eigenstates_2}). In the non degenerate subspace spanned by 
$|u^{(2)}\ra$ ``nothing evolves'', other than
a phase factor, but the initial states in the examples do not overlap with it.



\begin{thebibliography}{99}
\bibitem{review}E. Torrontegui, S. Ib\'a\~nez, S. Mart\'{i}nez-Garaot, M. Modugno, A. del Campo, D. Gu\'ery-Odelin, A. Ruschhaupt, X. Chen, and J. G. Muga, Adv. At. Mol. Opt. Phys. \textbf{62}, 117 (2013).
\bibitem{transport_2011} E. Torrontegui, S. Ib\'{a}\~{n}ez, X. Chen, A. Ruschhaupt, D. Gu\'{e}ry-Odelin and J. G. Muga, Phys. Rev. A \textbf{83}, 013415 (2011).
\bibitem{expansions_2010} J. G. Muga, X. Chen, S. Ib\'{a}\~{n}ez, I. Lizuain and A. Ruschhaupt, J. Phys. B: At. Mol. Opt. Phys., \textbf{43}, 085509 (2010).
\bibitem{2ls_2010} X. Chen, I. Lizuain, A. Ruschhaupt, D. Gu\'{e}ry-Odelin and J. G. Muga, Phys. Rev. Lett., \textbf{105}, 123003 (2010).
\bibitem{2ls_2011} X. Chen, E. Torrontegui and J. G. Muga, Phys. Rev. A \textbf{83}, 062116 (2011).
\bibitem{IP} S. Ib\'{a}\~{n}ez, X. Chen, E. Torrontegui, J. G. Muga and A. Ruschhaupt, Phys. Rev. Lett., \textbf{109}, 100403 (2012).
\bibitem{Molmer} T. Opatrn\'y and K. M\o lmer, New J. Phys. \textbf{16}, 015025 (2014).
\bibitem{Rice} M. Demirplak and S. A. Rice, J. Phys. Chem. A  \textbf{107}, 9937 (2003); J. Phys. Chem. B \textbf{109}, 6838 (2005); J. Chem. Phys., \textbf{129}, 154111 (2008).
\bibitem{Berry09} M. V. Berry, J. Phys. A \textbf{42}, 365303 (2009).
\bibitem{Rev_Longhi} S. Longhi, Laser \& Photon. Rev. \textbf{3}, 243 (2009).
\bibitem{Rev_Nolte} A. Szameit and S. Nolte, J. Phys. B: At. Mol. Opt. Phys., \textbf{43}, 163001 (2010).
\bibitem{Longhi_2011} S. Longhi, J. Phys. B: At. Mol. Opt. Phys. \textbf{44}, 051001 (2011).
\bibitem{Longhi_2008} M. Ornigotti, G. Della Valle, T. Toney Fernandez, A. Coppa, V Foglietti,
P. Laporta, and S. Longhi, J. Phys. B: At. Mol. Opt. Phys. \textbf{41},  085402 (2008).
\bibitem{Vitanov_2012} A. A. Rangelov and N. V. Vitanov, Phys. Rev. A \textbf{85}, 055803 (2012).

\bibitem{Tseng_2013} K.-H. Chien, C.-S. Yeih and S.-Y. Tseng, J. Lightw. Technol. \textbf{31}, 3387 (2013).

\bibitem{Fisher} M. P. A. Fisher, P. B. Weichman, G. Grinstein and D. S. Fisher, Phys. Rev. B \textbf{40}, 546 (1989).
\bibitem{Zoller} D. Jaksch, C. Bruder, J. I. Cirac, C. W. Gardiner and P. Zoller, Phys. Rev. Lett., \textbf{81}, 3108 (1998).


\bibitem{Mac} M. A. H. MacCallum in ``On Einstein's Path'' (199, Springer, New York, 1999) p. 299.  
\bibitem{Bambusi} D. Bambusi,  Nonlinearity \textbf{8}, 93 (1995). 
\bibitem{classical} J. R. Cary, Phys. Rep. \textbf{79}, 129 (1981). 
\bibitem{Mompart_2004} K. Eckert, M. Lewenstein, R. Corbalan, G. Birkl, W. Ertmer and J. Mompart, Phys. Rev. A \textbf{70}, 023606 (2004).
\bibitem{Lie} E. Torrontegui, S. Mart\'\i nez-Garaot and J. G. Muga, arXiv:1402.5695. 

\bibitem{Chen_3ls} X. Chen and J. G. Muga, Phys. Rev. A \textbf{86}, 033405 (2012).
\bibitem{rev_Shore} K. Bergmann, H. Theuer and B. Shore, Rev. Mod. Phys., \textbf{70}, 1003 (1998).
\bibitem{Kiselev} M. N. Kiselev, K. Kikoin, and M. B. Kenmoe, EPL \textbf{104}, 57004 (2013).
 \bibitem{Tseng_2012} T.-Y. Lin, F.-C. Hsiao, Y.-W. Jhang, C. Hu and S.-Y. Tseng, Opt. Express \textbf{20}, 24085 (2012).
\bibitem{Chen_2012} S.-Y. Tseng and X. Chen, Opt. Lett., \textbf{37}, 5118 (2012).

\bibitem{Dou} F. Q. Dou, L. B. Fu, and J. Liu, Phys. Rev. A \textbf{89}, 012123 (2014). 

\bibitem{patera} J. Patera, R. T. Sharp, P. Winternitz, and H. Zassenhaus, J. Math. Phys. \textbf{17}, 986 (1976). 
\bibitem{LR}H. R. Lewis and W. B. Riesenfeld, J. Math. Phys. \textbf{10} , 1458 (1969).

\end{thebibliography}
\end{document}